\documentclass[twocolumn,showpacs,preprintnumbers,amsmath,amssymb,prb]{revtex4}
\usepackage{graphicx}
\usepackage{dcolumn}
\usepackage{bm}
\begin{document}
\title{Thermodynamics and phase behavior of the lamellar Zwanzig model}
\author{Ludger Harnau$^{1,2}$, D. G. Rowan$^{3}$ and Jean-Pierre Hansen$^{3}$}
\address{
         $^{1}$Max-Planck-Institut f\"ur Metallforschung,  
         Heisenbergstr.\ 1, D-70569 Stuttgart (Germany) 
         \\
         $^{2}$Institut f\"ur Theoretische und Angewandte Physik, 
         Universit\"at Stuttgart, 
         Pfaffenwaldring 57, 
         D-70569 Stuttgart (Germany) 
         \\
         $^{3}$Department of Chemistry, University of Cambridge, Lensfield Road,
         Cambridge CB2 1EW (UK)
	 }
\date{\today}
\begin{abstract}
Binary mixtures of lamellar colloids represented by hard platelets are studied 
within a generalization of the Zwanzig model for rods, whereby the square cuboids 
can take only three orientations along the $x$, $y$ or $z$ axes. The free 
energy is calculated within Rosenfeld's ''Fundamental Measure Theory'' (FMT) adapted 
to the present model. In the one-component limit, the model exhibits the 
expected isotropic to nematic phase transition, which narrows as the aspect ratio 
$\zeta=L/D$ ($D$ is the width and $L$ the thickness of the platelets) increases. 
In the binary case the competition between nematic ordering and 
depletion-induced segregation leads to rich phase behaviour.
\end{abstract}

\oddsidemargin -12mm
\textwidth 188 mm
\maketitle

\section{Introduction}  
Suspensions of lamellar colloids, like clay dispersions, have received renewed 
experimental and theoretical attention lately, because of the expected rich 
phase behaviour, and the obvious geophysical and technological implications, 
in particular for oil drilling \cite{mait:00}. With increasing concentration 
most natural and synthetic clay dispersions undergo gelation \cite{mour:95}
which impedes the isotropic to nematic (I-N) transition expected from 
theoretical considerations \cite{onsa:49} and computer simulations 
\cite{eppe:84,veer:92} on entropic grounds. It came hence as a relief when a novel 
system of hard lamellar colloids, namely a dispersion of gibbsite platelets 
sterically stabilized by a grafted polymer layer, in toluene, was shown to exhibit
the expected I-N transition \cite{kooi:98}. The difference in behaviour between 
the gel-forming clays and the novel gibbsite system may lie in the long-range
Coulomb interactions present in the aqueous clay dispersions \cite{dijk:97}. 
Subsequently it was shown experimentally \cite{kooi:01}, theoretically 
\cite{wens:01}, and by simulation \cite{bate:99}, that polydispersity 
in the size of the platelets strongly affects the phase behaviour. In 
particular binary mixtures of thin and thick platelets lead to an unexpected 
isotropic-nematic density inversion \cite{kooi:01,wens:01}. Moreover, even 
for monodisperse systems, the phase diagram was shown to be quantitatively, 
and even qualitatively, sensitive to the aspect ratio (i.e., the thickness over 
diameter of platelets modeled as ''cut spheres'') \cite{veer:92}.

In this paper we address the problem of the phase behaviour of monodisperse
and bidisperse systems of hard platelets using a highly simplified model 
introduced some time ago by Zwanzig to study the I-N transition in systems 
of hard rods \cite{zwan:63}. In this model platelets or rods are represented
by square parallelepipeds (or cuboids) of thickness (length) $L$ and 
width $D$, which can take only three orientations, along the $x$, $y$ or $z$ 
directions, rather than a continuous range of orientations in space 
(cf. Fig.~\ref{fig1}). The aspect ratio $\zeta=L/D >1$ for rods, while 
$\zeta<1$ for platelets. The case $\zeta=1$ corresponds to a system of 
parallel cubes, for which analytic expressions of the first seven virial 
coefficients are known \cite{zwan:56}. Zwanzig's model may be considered 
as a highly coarse-grained version of the Onsager long-rod system with 
continuously varying orientation. The advantage is that virial coefficients 
beyond the second coefficient $B_2$ can be calculated in the limit 
$\zeta \gg 1$ thus providing a quantitative test of Onsager's
classical theory, which includes only $B_2$ \cite{onsa:49}.
In this paper we adapt Zwanzig's model to the platelet case ($\zeta<1$) 
to investigate the equation of state, the I-N transition, as well as 
binary mixtures of platelets of different size in the search of a 
possible, depletion-driven demixing transition \cite{rowa:02}.

\section{Model and virial expansion}
\begin{figure}
\hspace*{-1.5cm}
\includegraphics[width=0.45\linewidth]{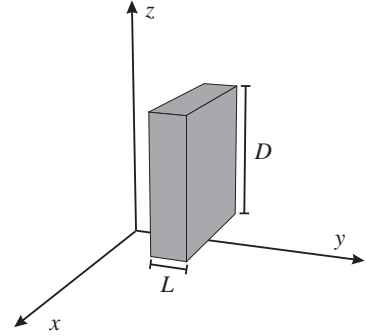}
\caption{In the model under consideration platelets are represented by 
square parallelepipeds of thickness $L$ and width $D$, which can take 
only three orientations, along the $x$, $y$ or $z$ directions. 
The number density of the centre of mass of the platelet displayed 
in the figure is denoted by $\rho_y$.}
\label{fig1}
\end{figure}

Following up on Zwanzig's original idea for rods, we consider a system of 
$N$ hard square cuboids, of dimension $D \times D \times L$, as shown in 
Fig.~\ref{fig1}, with $L<D$ ($\zeta<1$). 
The normals to these platelets are restricted 
to point along one of the Cartesian axes, and platelets of a given orientation 
are considered as belonging to one of three ``species''; the numbers of 
platelets belonging to each species are denoted by $N_x$, $N_y$ and $N_z$, 
with the total number of platelets equal to $N=N_x+N_y+N_z$. $N$ is fixed, 
but the $N_\alpha$ fluctuate due to thermal reorientations of the platelets. 
The mean numbers of platelets per unit volume will be denoted by
$\rho_\beta=<N_\beta>/V$. In an isotropic phase, $\rho_x=\rho_y=\rho_z=\rho/3$, 
where $\rho=N/V$.

The virial expansion of the excess free energy per unit volume, 
in powers of the partial densities $\rho_\beta$, is:
\begin{equation} \label{eq1}
f^{ex}(\{\rho_\beta\})=\frac{F^{ex}}{Vk_BT}=\sum\limits_{i=2}^\infty
\frac{1}{i-1}B_i(\beta_1, ... , \beta_i)\rho_{\beta_1}\cdot\cdot\cdot
\rho_{\beta_i}\,,
\end{equation}
where the indices $\beta_1$, $\beta_2$, ... sum over \mbox{$x$, $y$, and $z$.} 
$B_2(\beta_1,\beta_2)$ is the second virial coefficient between platelets 
of species $\beta_1$ and $\beta_2$:
\begin{equation} \label{eq2}
B_2(\beta_1,\beta_2)=-\frac{1}{2}\int d{\bf r}\, f_{\beta_1,\beta_2}({\bf r})
\end{equation}
where for cuboids, the Mayer functions reduce to products of Heaviside 
step functions:
\begin{eqnarray} \label{eq3}
f_{\beta_1,\beta_2}({\bf r})&=&-\prod_{\alpha=1}^3
\Theta\left(\frac{1}{2}\left(S_{\alpha,\beta_1}+S_{\alpha,\beta_2}
\right)-|r_{\alpha}|\right)\,.
\end{eqnarray}
In Eq.~(\ref{eq3}), $S_{\alpha,\beta}$ is the spatial extent of plates 
of species (orientation) $\beta$ in the direction $\alpha=x,y,z$:
\begin{equation} \label{eq4}
S_{\alpha,\beta}=D+(L-D)\delta_{\alpha,\beta}\,,
\end{equation}
while $|r_{\alpha}|$ is the absolute value of the $\alpha$-component of 
the vector ${\bf r}_{\alpha}$. For obvious symmetry reasons, there are only 
two independent $B_2(\beta_1,\beta_2)$, namely 
$B_2(x,x)=B_2(y,y)=B_2(z,z)$ (parallel platelets) and 
$B_2(x,y)=B_2(x,z)=B_2(y,z)$ (orthogonal platelets). The corresponding integrals 
(\ref{eq2}) are easily calculated, to yield:
\begin{equation} \label{eq5}
B_2(x,x)=4LD^2\,;\hspace{0.2cm}B_2(x,y)=D(L+D)^2\,.
\end{equation}
Similarly, there are 3 independent third virial coefficients, namely 
$B_3(x,x,x)$, $B_3(x,x,y)$ and  $B_3(x,y,z)$. These are calculated in
Appendix A, together with independent fourth virial coefficients 
(the latter only in the limits $L\to 0$, $D\to 0$, and $D=L$).
Gathering results, the reduced excess 
free energy density may be cast in the form: 
\begin{eqnarray}  \label{eq6}
\lefteqn{f^{ex}=}\nonumber
\\&&B_2(x,x)\left(\rho_x^2+\rho_y^2+\rho_z^2\right)\nonumber
\\&&+B_2(x,y)\left(\rho_x\rho_y+\rho_x\rho_z+\rho_y\rho_z\right)\nonumber
\\&&+\frac{1}{2}B_3(x,x,x)\left(\rho_x^3+\rho_y^3+\rho_z^3\right)
+\frac{1}{2}B_3(x,y,z)\rho_x\rho_y\rho_z\nonumber
\\&&+\frac{1}{2}B_3(x,x,y)\left[\rho_x^2(\rho_y+\rho_z)+\rho_y^2(\rho_x+\rho_z)
+\rho_z^2(\rho_x+\rho_y)\right]\nonumber
\\&&+O(\rho^4)\,,
\end{eqnarray}
and the resulting equation of state reads:
\begin{eqnarray}  \label{eq7}
\lefteqn{\frac{P}{k_BT}=}\nonumber
\\&&\rho+B_2(x,x)\left(\rho_x^2+\rho_y^2+\rho_z^2\right)\nonumber
\\&&+B_2(x,y)\left(\rho_x\rho_y+\rho_x\rho_z+\rho_y\rho_z\right)\nonumber
\\&&+B_3(x,x,x)\left(\rho_x^3+\rho_y^3+\rho_z^3\right)
+B_3(x,y,z)\rho_x\rho_y\rho_z\nonumber
\\&&+B_3(x,x,y)\left[\rho_x^2(\rho_y+\rho_z)+\rho_y^2(\rho_x+\rho_z)
+\rho_z^2(\rho_x+\rho_y)\right]\nonumber
\\&&+O(\rho^4)\,.
\end{eqnarray}
It is instructive to consider the isotropic (low density) phase first, 
where $\rho_x=\rho_y=\rho_z=\rho/3$. Introducing the total (or 
orientation-averaged) virial coefficients:
\begin{eqnarray}  \label{eq8a}
B_2&=&\frac{1}{3}\left[B_2(x,x)+2B_2(x,y)\right]\,,
\\B_3&=&\frac{1}{9}\left[B_3(x,x,x)+6B_3(x,x,y)+2B_3(x,y,z)\right]\,, \label{eq8b}
\end{eqnarray}
the equation of state of the isotropic phase reduces to the usual form:
\begin{equation}  \label{eq9}
\frac{P}{k_BT}=\rho+B_2\rho^2+B_3\rho^3+O(\rho^4)\,.
\end{equation}
The two important dimensionless parameters in the problem are the size 
ratio $\zeta=L/D$, and the reduced density (or packing fraction) 
$\rho^\star=\rho D^2L$. Substituting the results from Appendix A, 
the dimensionless equation of state $P/(k_BT\rho)$ may be expressed in terms 
of these two variables as:
\begin{eqnarray}  \label{eq10}
&&\frac{P}{k_BT\rho}=1+\frac{2}{3}\left(\frac{1}{\zeta}+4+\zeta\right)
\rho^\star\nonumber
\\&&+\frac{1}{27}\left(\frac{2}{\zeta^2}+\frac{48}{\zeta}+141+52\zeta\right)
\rho^{\star 2}+O(\rho^{\star 3})\,.
\end{eqnarray}
It is worthwhile to examine three limiting cases, including contributions up 
to the fourth virial coefficient.

a) Long rods, corresponding to the limit $\zeta \to \infty$; in that case the 
relevant variable is the Onsager parameter, or effective packing fraction 
$\eta=\rho^\star\zeta=\rho DL^2$, in terms of which:
\begin{eqnarray}  \label{eq11}
\frac{P^{(\rm rods)}}{k_BT\rho}&=&1+\frac{2}{3}\eta-\frac{16}{243}\eta^3+O(\eta^4)\,,
\end{eqnarray}
i.e., the third virial contribution $\sim \eta^2$ is identically zero, in 
agreement with Onsager's conjecture for rods with continuous orientations 
\cite{onsa:49}, and with Zwanzig's result \cite{zwan:63}.

b) Thin platelets, corresponding to the limit $\zeta\to 0$; in that case 
the relevant variable is the effective packing fraction 
$\eta=\rho^\star/\zeta=\rho D^3$,
\begin{eqnarray}  \label{eq12}
\frac{P^{(\rm platelets)}}{k_BT\rho}&=&1+\frac{2}{3}\eta+\frac{2}{27}\eta^2-
\frac{32}{243}\eta^3+O(\eta^4)\,,
\end{eqnarray}
In this case the third virial coefficient $B_3$ is non-zero, nor is $B_4$, in 
agreement with numerical findings for continuously rotating disc-shaped 
platelets \cite{eppe:84}. At this stage there is no indication of convergence 
of the virial series at high packing fraction, but some plausibility 
arguments will be put forward in the following section to justify 
truncation of the virial series (\ref{eq12}) after the $O(\eta^2)$ 
(third virial) term for infinitely thin platelets.

c) Parallel cubes, corresponding to $\zeta=1$; the relevant dimensionless 
variable is the packing fraction $\eta=\rho^\star$, and the first few terms 
in the virial series read \cite{zwan:56}:
\begin{eqnarray}  \label{eq13}
\frac{P^{(\rm cubes)}}{k_BT\rho}&=&1+4\eta+9\eta^2+\frac{34}{3}\eta^3+
O(\eta^4)\,.
\end{eqnarray}

In Figure 2 the low density equation of state is shown for several aspect 
ratios $0\le\zeta\le 1$, and compared to available simulation data for 
hard cubes \cite{swol:87} and for infinitely thin discs of the same area 
($\pi R^2=D^2$), but, with continuously varying orientations 
\cite{eppe:84,dijk:97}. As expected, the pressure increases rapidly with 
$\zeta$ for any reduced density $\rho D^3$. The results for continuously 
rotating discs and for squares with discrete orientations are surprisingly 
close.
\begin{figure}
\vspace*{-1.0cm}
\includegraphics[width=1\linewidth]{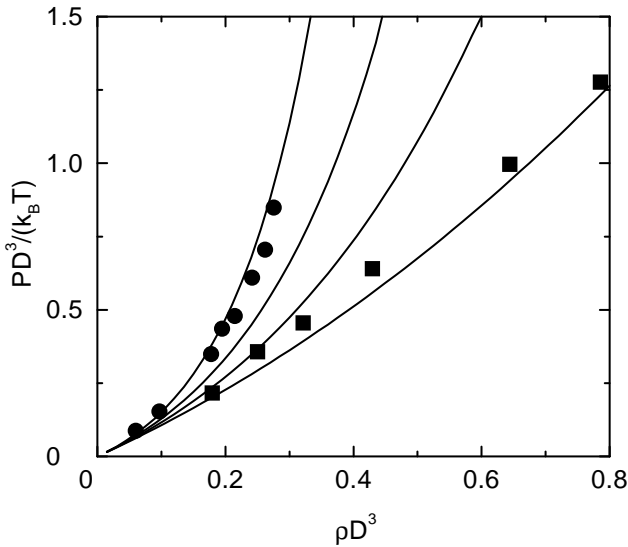}
\vspace*{-4.5cm}
\caption{Low density equation of state of a monodisperse, isotropic fluid consisting of
rectangular platelets of surface size $D \times D$ and thickness $L$ for various aspect
ratios: $\zeta=L/D=0, 0.3, 0.6, 1$ (solid lines from bottom to top).
The symbols represent computer simulation data for parallel hard cubes
(circles, $D=L$) \protect\cite{swol:87} and thin circular platelets
(squares, $L\to 0$) \protect\cite{eppe:84,dijk:97} with continuous
allowed orientations.}
\label{fig2}
\end{figure}

\section{Fundamental measure theory for hard platelets}
In view of the uncertain convergence of the virial series of the equation 
of state of hard platelets at high densities ($\eta \ge 1$), it is 
important to seek an approximate resummation of the series, similar 
to scaled particle theory for hard spheres \cite{reis:59}. This is 
most conveniently achieved within the framework of Rosenfeld's FMT 
\cite{rose:88}, by generalizing Cuesta's work for hard parallel cubes 
\cite{cues:96} to the case of Zwanzig's model for hard platelets. The 
second and third virial coefficients calculated in the preceding section
provide key input into the theory. Since the free energy of bidisperse 
platelets will be required to construct complete phase diagrams in 
section V, the FMT is formulated here for a binary mixture of platelets
of size $D_i\times D_i\times L_i$ ($i=1,2$). It is convenient to consider 
the more general case of inhomogeneous mixtures, characterized by the 
local densities $\rho^{(i)}_\beta({\bf r})$ of the centres of mass of 
platelets of species $i$, with orientations along $\beta=x,y,z$.

Under the influence of external potentials $V^{(i)}_\beta({\bf r})$, 
the equilibrium density profile of the mixture minimize the grand potential 
functional:
\begin{eqnarray} \label{eq14}
\Omega[\rho^{(i)}_\beta({\bf r})]&=&\sum_{i=1}^2\sum_{\beta}\int d{\bf r}\,
\rho^{(i)}_\beta({\bf r})
\left[k_BT\left(\ln[\Lambda_i^3\rho^{(i)}_\beta({\bf r})]-1\right)\right.
\nonumber
\\&&-\left.\mu_i+ V^{(i)}_\beta({\bf r})\right]+F_{ex}[\rho^{(i)}_\beta({\bf r})]\,,
\end{eqnarray}
where the $\mu_i$ are the chemical potentials of the two species (irrespective
of their orientation), while the $\Lambda_i$ are the thermal de Broglie 
wavelengths. 
The basic assumption of FMT is to postulate
the following form of the excess free energy functional:
\begin{equation} \label{eq15}
F_{ex}[\rho^{(i)}_\beta({\bf r})]=k_BT\int d{\bf r}\,\Phi[n_l({\bf r})]\,,
\end{equation}
where the reduced  excess free energy density $\Phi$ is a function of a set of 
weighted densities
\begin{equation} \label{eq16}
n_l({\bf r})=\sum_{i=1}^2\sum_\beta\int d{\bf r}_1\rho^{(i)}_\beta({\bf r}_1)
\omega_{l,\beta}^{(i)}({\bf r}-{\bf r}_1)\,.
\end{equation}
The weight functions $\omega_{l,\beta}^{(i)}({\bf r})$ are determined by
expressing the Fourier transforms of the Mayer functions 
$f^{(i,j)}_{\beta_1,\beta_2}$ as sums of products of functions associated 
with single platelets. In the case of mixtures, equation (\ref{eq3}) is 
generalized as:
\begin{eqnarray} \label{eq17}
\lefteqn{f^{(i,j)}_{\beta_1,\beta_2}({\bf r}_1,{\bf r}_2)=}\nonumber
\\&&-\prod_{\alpha=1}^3
\Theta\left(\frac{1}{2}\left(S_{\alpha,\beta_1}^{(i)}+S_{\alpha,\beta_2}^{(j)}
\right)-|r_{\alpha,1}-r_{\alpha,2}|\right)\,,
\end{eqnarray}
where \mbox{$S_{\alpha,\beta}^{(i)}=D_i+(L_i-D_i)\delta_{\alpha,\beta}$}.
The Fourier transform of the Mayer function (\ref{eq17}) may be decomposed 
according to:
\begin{eqnarray} \label{eq18}
&&\hspace{-0.5cm} f^{(i,j)}_{\beta_1,\beta_2}({\bf q})=\int\limits^\infty_{-\infty}d {\bf r}\,
e^{i{\bf q}\cdot{\bf r}}f^{(i,j)}_{\beta_1,\beta_2}({\bf r})\nonumber
\\&=&\omega^{(i)}_{0,\beta_1}({\bf q})\omega^{(j)}_{3,\beta_2}({\bf q})
+\omega^{(i)}_{3,\beta_1}({\bf q})\omega^{(j)}_{0,\beta_2}({\bf q})\nonumber
\\&&+\mbox{\boldmath$\omega$}^{(i)}_{1,\beta_1}({\bf q})\cdot
\mbox{\boldmath$\omega$}^{(j)}_{2,\beta_2}({\bf q})
+\mbox{\boldmath$\omega$}^{(i)}_{2,\beta_1}({\bf q})\cdot
\mbox{\boldmath$\omega$}^{(j)}_{1,\beta_2}({\bf q})\,,
\end{eqnarray}
where the scalar and vectorial weights ($0 \le l \le 3$) are defined by:
\begin{eqnarray}
\omega^{(i)}_{0,\beta}({\bf q})&=&
b^{(i)}_{x,\beta}(q_x)b^{(i)}_{y,\beta}(q_y)b^{(i)}_{z,\beta}(q_z)\,, \label{eq19}
\\\nonumber\\\mbox{\boldmath$\omega$}^{(i)}_{1,\beta}({\bf q})&=&
\left(
\begin{array}{l}
a^{(i)}_{x,\beta}(q_x)b^{(i)}_{y,\beta}(q_y)b^{(i)}_{z,\beta}(q_z)\\
\\
b^{(i)}_{x,\beta}(q_x)a^{(i)}_{y,\beta}(q_y)b^{(i)}_{z,\beta}(q_z)\\
\\
b^{(i)}_{x,\beta}(q_x)b^{(i)}_{y,\beta}(q_y)a^{(i)}_{z,\beta}(q_z)
\end{array}\right)\,,\label{eq20}
\\\nonumber\\\mbox{\boldmath$\omega$}^{(i)}_{2,\beta}({\bf q})&=&
\left(
\begin{array}{l}
b^{(i)}_{x,\beta}(q_x)a^{(i)}_{y,\beta}(q_y)a^{(i)}_{z,\beta}(q_z)\\
\\
a^{(i)}_{x,\beta}(q_x)b^{(i)}_{y,\beta}(q_y)a^{(i)}_{z,\beta}(q_z)\\
\\
a^{(i)}_{x,\beta}(q_x)a^{(i)}_{y,\beta}(q_y)b^{(i)}_{z,\beta}(q_z)
\end{array}\right)\,,\label{eq21}
\\\nonumber\\\omega^{(i)}_{3,\beta}({\bf q})&=&
a^{(i)}_{x,\beta}(q_x)a^{(i)}_{y,\beta}(q_y)a^{(i)}_{z,\beta}(q_z)\,, \label{eq22}
\end{eqnarray}
and:
\begin{eqnarray}  \label{eq23}
a^{(i)}_{\alpha,\beta}(q_\alpha)&=&\frac{2}{q_\alpha}
\sin\left(\frac{S^{(i)}_{\alpha,\beta}q_\alpha}{2}\right)\,,
\end{eqnarray}
\begin{eqnarray}  \label{eq24}
b^{(i)}_{\alpha,\beta}(q_\alpha)=\cos\left(\frac{S^{(i)}_{\alpha,\beta}q_\alpha}{2}\right)\,.
\end{eqnarray}
Substitution of the Fourier transforms of the weight functions 
(\ref{eq19})-(\ref{eq22}) into equation (\ref{eq16}) yields two scalar 
and two vectorial weighted densities $n_l({\bf r})$. Letting  
$n_l({\bf r})={\bf n}_l({\bf r})\cdot{\bf u}$, $l=1,2$, with ${\bf u}=(1,1,1)$, 
one can construct the function $\Phi$ from the following linear combination 
of scalar terms, each of dimension (volume)$^{-1}$:
\begin{eqnarray}   \label{eq25}
\lefteqn{\Phi[n_l({\bf r})]=}\nonumber
\\&&c_0n_0({\bf r})+c_1n_1({\bf r})n_2({\bf r})+
c_2{\bf n}_1({\bf r})\cdot{\bf n}_2({\bf r})
+c_3n_2^3({\bf r})\nonumber
\\&&+c_4n_2{\bf n}_2({\bf r})\cdot{\bf n}_2({\bf r})+
c_5{\bf n}_2({\bf r})\cdot{\bf n}_2({\bf r})\cdot{\bf n}_2({\bf r})\,,
\end{eqnarray}
where ${\bf n}\cdot{\bf n}\cdot{\bf n}=n_x^3+n_y^3+n_z^3$. The 6 coefficients
$c_i$ ($0 \le i \le 5$) are analytical functions of the only dimensionless
weighted density, namely $n_3$. These functions are determined, within integration 
constants, by the scaled particle condition \cite{reis:59}, which results in the 
following differential equation for $\Phi$ \cite{rose:88}:
\begin{equation} \label{eq26}
-\Phi+\sum\limits_\beta\rho_\beta\frac{\partial \Phi}{\partial \rho_\beta}
+n_0=\frac{\partial \Phi}{\partial n_3}\,.
\end{equation}
Substitution of (\ref{eq25}) into (\ref{eq26}) yields, upon integration:
\begin{eqnarray} \label{eq27}
c_0&=&-\ln(1-n_3)\,,  c_1=\frac{d_1}{1-n_3}\,, c_2=\frac{d_2}{1-n_3}\,,\nonumber
\\c_3&=&\frac{d_3}{(1-n_3)^2}\,, c_4=\frac{d_4}{(1-n_3)^2}\,, c_5=\frac{d_5}{(1-n_3)^2}\,.
\end{eqnarray}
The remaining constants $d_i$ ($1 \le i \le 5$) are determined by identifying
the low density expansion of the excess free energy, defined by Eqs.~(\ref{eq15}), 
(\ref{eq25}) and (\ref{eq27}) for the homogeneous fluid 
(i.e., $\rho_\beta^{(i)}({\bf r})=\rho_\beta^{(i)}$) with the third order virial 
expansion (\ref{eq1}) of the same free energy; this leads to 
\begin{eqnarray} \label{eq28}
d_1=0\,, d_2=1\,, d_3=\frac{1}{6}\,, d_4=-\frac{1}{2}\,, d_5=\frac{1}{3}\,.
\end{eqnarray}
The final expression for the function $\Phi$ then reads:
\begin{eqnarray} \label{eq29}
\Phi=-n_0\ln(1-n_3)+\frac{{\bf n}_1\cdot{\bf n}_2}{1-n_3}+
\frac{n_{2,x}n_{2,y}n_{2,z}}{(1-n_3)^2}\,,
\end{eqnarray}
where $n_{2,\alpha}$ is the projection of the vector ${\bf n}_2$ in the $\alpha$
direction.
The FMT expression for the free energy functional is now completely specified.
In the limit of parallel hard cubes ($L_i=D_i$) the excess free energy functional 
reduces to the result of Ref.~\cite{cues:96}. 

In the limit of a homogeneous 
bulk fluid the equilibrium density profiles are constant, and the
weighted densities take the explicit expressions:
\begin{eqnarray}  \label{eq30}
n_0&=&\sum_{i=1}^2\left(\rho^{(i)}_x+\rho^{(i)}_y+\rho^{(i)}_z\right)\,, 
\\\nonumber\\{\bf n}_1&=&\sum_{i=1}^2
\left(
\begin{array}{l}
L_i\rho^{(i)}_x+D_i\rho^{(i)}_y+D_i\rho^{(i)}_z\\\\
D_i\rho^{(i)}_x+L_i\rho^{(i)}_y+D_i\rho^{(i)}_z\\\\
D_i\rho^{(i)}_x+D_i\rho^{(i)}_y+L_i\rho^{(i)}_z
\end{array}\right)\,,\label{eq31}
\\\nonumber\\{\bf n}_2&=&\sum_{i=1}^2
\left(
\begin{array}{l}
D_i\rho^{(i)}_x+L_i\rho^{(i)}_y+L_i\rho^{(i)}_z\\\\
L_i\rho^{(i)}_x+D_i\rho^{(i)}_y+L_i\rho^{(i)}_z\\\\
L_i\rho^{(i)}_x+L_i\rho^{(i)}_y+D_i\rho^{(i)}_z
\end{array}\right)D_i\,,\label{eq32}
\\\nonumber\\n_3&=&\sum_{i=1}^2L_iD_i^2\left(\rho^{(i)}_x+\rho^{(i)}_y+\rho^{(i)}_z\right)\,, 
\label{eq33}
\end{eqnarray}
The weight functions may be interpreted in terms of the following geometric 
characteristics of the particles: their mean diameter  $(L_i+2D_i)/3$, their
surface $2D_i(D_i+2L_i)$, and their volume $L_iD_i^2$.

In the limit of an isotropic ($\rho_x=\rho_y=\rho_z=\rho/3$) one-component 
system of platelets, the equation-of-state derived from the free energy 
density (\ref{eq29}) takes the following form (with $\eta=\rho D^3$, the 
effective packing fraction, and $\zeta=L/D$): 
\begin{eqnarray} \label{eq34}
P^\star=\frac{PD^3}{k_BT}&=&\eta+\frac{\zeta\eta^2}{1-\eta\zeta}+
\frac{(2+5\zeta)\eta^2}{3(1-\eta\zeta)^2}\nonumber
\\&&+\frac{2(1+2\zeta)^3\eta^3}{27(1-\eta\zeta)^3}\,.
\end{eqnarray}
Interestingly, in the $\zeta \to 0$ limit (infinitely thin platelets) 
the FMT equation-of-state reduces to the virial series (\ref{eq12}), 
truncated after third order. This is reminiscent of the equation of state 
derived for circular platelets (discs), with continuously varying 
orientation, from PRISM (polymer reference interaction site model), which 
also reduces to a three term series (although the coefficients of $\eta^2$ 
and $\eta^3$ differ slightly from the exact virial coefficient for that 
case) \cite{harn:01}. These two observations prompt the conjecture that 
a three term virial series may be quantitatively accurate for infinitely thin 
platelets (with discrete or continuous orientations) beyond the low density regime.
 
For finite thickness, however ($\zeta >0$), the three-term virial series and 
the FMT equation-of-state (\ref{eq34}) deviate increasingly as $\zeta$ and 
$\eta$ increase. Thus, for $\eta=1$, $P^\star_1$, as calculated from 
Eq.~(\ref{eq10}) and $P^\star_2$, derived from Eq.~(\ref{eq34}),
take the values $P^\star_1=2.246$ and 
$P^\star_2=2.315$ for $\zeta=0.1$, while $P^\star_1=2.88$ and 
$P^\star_2=3.209$ for $\zeta=0.2$; for $\zeta=0.5$, the three-term 
virial series underestimates the pressure by more than a factor of $2$ 
($P^\star_1=5.676$ and $P^\star_2=12.74$) for $\eta=1$.

\section{Nematic ordering of platelets}
As a direct application of the FMT free energy (\ref{eq29}) derived in the 
preceding section, we now consider the possibility of an isotopic to nematic
(I-N) transition in a monodisperse fluid of platelets, as a function of the 
the aspect ratio $\zeta=L/D$. The broken symmetry in the nematic phase may be 
characterized by the order parameter:
\begin{equation} \label{eq35}
s=\frac{\rho_z-(\rho_x+\rho_y)/2}{\rho}\,,
\end{equation}
which varies between $0$ (isotropic phase) and $1$ (when all platelets 
are oriented along the $z$-axis). In the nematic phase $\rho_x=\rho_y$, 
and the independent variables are $\rho$ and $s$. For given values of 
$\rho$, the one-component version of the grand potential (\ref{eq14}) 
(with $V_\beta({\bf r})=0$, and $\rho_\beta({\bf r})=\rho_\beta$ and 
$F_{ex}$ given by (\ref{eq15}) and (\ref{eq29})) is minimized with respect to 
the order parameter $s$. If the FMT excess free energy (\ref{eq29}) is 
replaced by the virial expansion (\ref{eq6}), truncated after third order 
(which is a good approximation for thin platelets, $\zeta \ll 1$), the 
following Euler-Langrange equation determines $s$ as a 
\mbox{function of $\rho$:}
\begin{eqnarray}  
&&\ln\frac{1+2s}{1-s}=2D(L-D)^2\rho s\nonumber
\\&&+\frac{D^3}{3}(D+8L)(L-D)^2\rho^2 s
+\frac{D^3}{3}(L-D)^3\rho^2s^2\,. \label{eq36}
\end{eqnarray}
For low densities $\rho$, only the isotropic phase ($s=0$) is stable. At
a critical density $\rho_c$, a second, non-zero solution of Eq.~(\ref{eq36})
emerges, which eventually will lead to a stable nematic phase, when 
the free energy associated with $s \neq 0$ drops below that of the 
isotropic phase. The non-zero root $s$ requires a numerical solution 
of Eq.~(\ref{eq36}). A simple estimate of the critical density required to 
obtain a non-zero value of $s$ is obtained from the small $s$ expansion 
of the l.h.s of Eq.~(\ref{eq36}). If only terms to linear order in $\rho$ and 
$s$ are retained, the bifurcation density is found to be:
\begin{eqnarray}  \label{eq37}
\rho^{(c)}&\approx&\frac{3}{2D(L-D)^2}\,.
\end{eqnarray}
Including quadratic terms leads to $\rho_c D^3=1.243$ in the $\zeta \to 0$
limit. Because of the restricted number of allowed orientations, the
I-N transition takes place at relatively low density compared 
to the  predictions  of simulations for thin circular platelets with 
continuous orientation \cite{eppe:84,dijk:97}. This point has already 
been discussed in the case of rod-like particles ($\zeta > 1$) 
\cite{stra:72}. The estimate (\ref{eq37}) predicts that $\rho_c$ 
{\em increases} upon increasing the aspect ratio $\zeta$ ($<1$),
in agreement with recent experimental data \cite{kooi:01} and 
theoretical calculations \cite{wens:01}.

\begin{figure}
\vspace*{-1.0cm}
\includegraphics[width=1\linewidth]{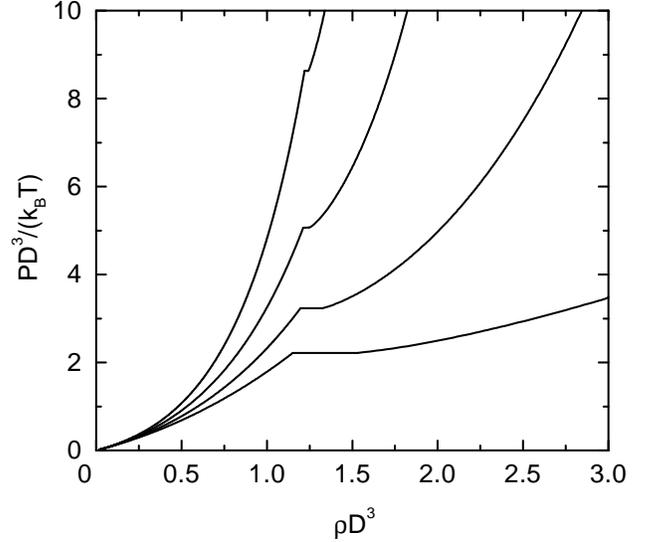}
\vspace*{-4.5cm}
\caption{Equation of state of a monodisperse fluid consisting of
rectangular platelets of surface size $D \times D$ and thickness $L$ for 
various aspect ratios: $\zeta=L/D=0.01, 0.1, 0.2, 0.3$ (from bottom to top). 
The horizontal lines are tie-lines illustrating isotopic-nematic 
coexistence.}
\label{fig3}
\end{figure}

For $\zeta>0$, quantitatively more reliable phase diagrams may be derived 
by minimizing the grand potential (\ref{eq14}), with the FMT excess free 
energy density (\ref{eq29}), with respect to the order parameter $s$
for each density $\rho$. The resulting equations of state in the isotropic 
($s=0$) and nematic ($s\neq 0$) phases are plotted versus $\rho$ in 
Fig.~\ref{fig3}, together with the horizontal tie-lines connecting
the two phases, for several aspect ratios $\zeta$. The value of the effective
packing fraction $\eta_I=\rho_I D^3$ and $\eta_N=\rho_N D^3$ of the coexisting 
phase are seen to be rather insensitive to the aspect ratio $\zeta$, 
while the first order transition narrows as $\zeta$ increases, i.e., 
$\triangle \eta=\eta_N-\eta_I$ \mbox{decreases with increasing $\zeta$.}
The reduced pressure $P^\star= PD^3/(k_BT)$, however, increases rapidly with 
$\zeta$, as one might expect. 

In order to study the possible onset of an uniaxial nematic phase ($\rho_x=\rho_y$), 
we rewrite the Euler-Lagrange equations in terms of $\rho$ and $s$:
\begin{eqnarray}  
\ln\frac{1+2s}{1-s}&=&\frac{2D(L-D)^2\rho s}{1-D^2L\rho}\nonumber
\\&+&\frac{D^3(L-D)^2\rho^2s}{3(1-D^2L\rho)^2}
(D+2L+[L-D]s)\,. \label{eq37a}
\end{eqnarray}
If only terms to cubic order in $\rho$ are retained, 
Eq.~(\ref{eq37a}) reduces to Eq.~(\ref{eq36}). The bifurcation density follows 
from a low-$s$ expansion of Eq.~(\ref{eq37a}):
\begin{eqnarray} \label{eq37b}
\rho^{(c)}\approx \frac{G}{1+GD^2L}\,,
\end{eqnarray}
with 
\begin{eqnarray} \label{eq37c}
G&=&-\frac{3}{D^2(D+2L)}\nonumber
\\&&+\frac{3}{D^2}\sqrt{\frac{1}{(D+2L)^2}+
\frac{D}{(D+2L)(L-D)^2}}\,.
\end{eqnarray}
Thus, for $\zeta=L/D=0.2, 0.4, 0.6, 0.8$, $\rho^{(c)}$, as calculated from 
Eq.~(\ref{eq37c}), takes the values $\rho^{(c)}D^3=1.259, 1.229, 1.165, 1.085$,
which are rather insensitive to the aspect ratio $\zeta$.

\section{Phase diagrams of binary platelet mixtures}
We next consider the richer case of highly asymmetric binary mixtures of
large and small platelets. Polydispersity, both in diameters and in thickness,
is intrinsic to the experimental gibbsite samples which have been recently
studied \cite{kooi:98,kooi:01}.
In the case of mixtures of spherical colloidal particles large asymmetry may lead 
to depletion-induced phase separation \cite{dijk:99}. In the case of platelets
depletion-induced segregation competes with nematic ordering, as is also
the case for mixtures of long and short rods \cite{voer:92}, or of thin and thick
rods \cite{roij:98}. Recent preliminary work on mixtures of continuously rotating
thin circular platelets ($\zeta=0$) suggests that depletion-induced fractionation
is strongest when the large platelets are nematically ordered \cite{rowa:02}. 
Fractionation was also predicted for thin and thick circular platelets
\cite{wens:01}, on the basis of Parsons' heuristic extension 
of Onsager's theory \cite{onsa:49,pars:79}.

\begin{figure}
\vspace*{-1.0cm}
\includegraphics[width=1\linewidth]{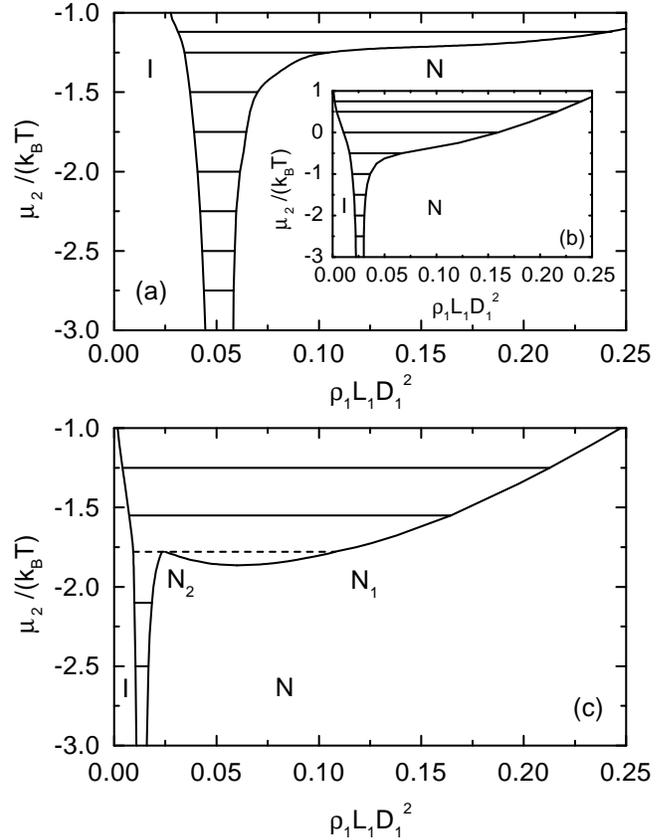}
\vspace*{-1cm}
\caption{Phase diagram of a fluid consisting of a binary mixture of small
platelets [$L_2/D_2=0.1$] and large platelets 
[(a): $D_1/D_2=10$, $L_1/D_1=0.04$;
(b): $D_1/D_2=5$, $L_1/D_1=0.04$;
(c): $D_1/D_2=10$, $L_1/D_1=0.01$] as a function of the chemical
potential of the small platelets $\mu_2$ and the density of the large
platelets $\rho_1$. The straight solid lines are tie-lines illustrating
isotropic-nematic (I-N) coexistence. The dashed line in (c) marks coexistence 
of an isotropic phase (I) with two nematic phases (N$_1$ and N$_2$) of 
different compositions.}
\label{fig4}
\end{figure}

\begin{figure}
\vspace*{-1.0cm}
\includegraphics[width=1\linewidth]{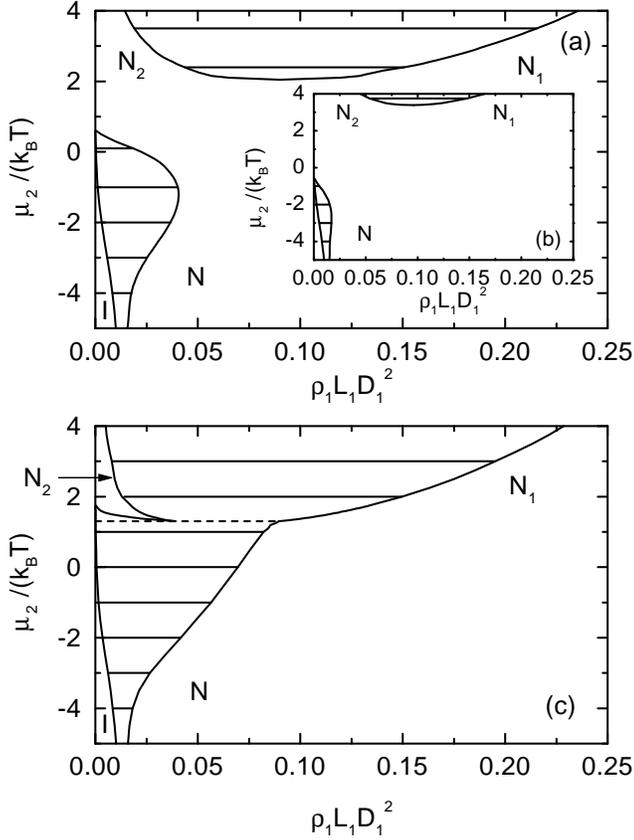}
\vspace*{-1cm}
\caption{Phase diagram of a fluid consisting of a binary mixture of large
thin platelets [$L_1/D_1=0.01$] and small platelets
[(a): $D_1/D_2=2$, $L_2/D_2=0.16$;
(b): $D_1/D_2=1.66$, $L_2/D_2=0.13$;
(c): $D_1/D_2=2$, $L_2/D_2=0.2$] as a function of the chemical
potential of the small platelets $\mu_2$ and the density of the large
platelets $\rho_1$. The straight solid lines are tie-lines illustrating
isotropic-nematic (I-N) and nematic-nematic (N$_1$-N$_2$) coexistence,
respectively. The dashed line in (c) marks coexistence
of an isotropic phase (I) with two nematic phases (N$_1$ and N$_2$).}
\label{fig5}
\end{figure}

In this section we consider binary mixtures of square platelets, using the
FMT excess free energy (\ref{eq29}). Various phase diagrams constructed 
as function of the chemical potential $\mu_2$ of the small platelets  and 
the number density $\rho_1$ of the large platelets ($D_1>D_2$) are shown 
in Figs.~\ref{fig4}, \ref{fig5}. The tie-lines are horizontal because of 
equality of $\mu_2$ of the coexisting phases.
For large negative chemical potential $\mu_2$, the systems exhibit a 
first oder I-N transition and the density gap at the I-N transition broadens 
with increasing $\mu_2$ in agreement with our previous calculations for 
binary mixtures of infinitely thin platelets \cite{harn:02}. Upon increasing 
$\mu_2$, the phase diagrams presented in  Figs.~\ref{fig4}, \ref{fig5}
differ qualitatively, and it is worthwhile to distinguish the following 
cases.

Fig.~\ref{fig4} (a), (b): The width of the I-N transition broadens 
continuously with increasing $\mu_2$. At high densities in the nematic phase,
the small platelets are nearly excluded and hence the large platelets are
in equilibrium with a reservoir of small platelets. This implies that 
the pressure of the nematic phase is equal to that of a reservoir of 
small platelets at a chemical potential $\mu_2$.

Fig.~\ref{fig4} (c): Decreasing the aspect ratio $\zeta_1=L_1/D_1$ 
(as compared to Fig.~\ref{fig4} (a), (b)) at fixed $L_2$, $D_2$
leads to a smaller width of the I-N transition and a shift of the 
I-N transition density to smaller values of $\rho_1$ as already 
discussed in the case of monodisperse platelet fluids
(see Eq.~(\ref{eq37})).
The I-N transition ends
in an I-N$_1$-N$_2$ triple point, at which two nematic phases 
(N$_1$, N$_2$) coexist with an isotropic phase. Above the triple point there 
is coexistence between a low-density isotropic and a high-density nematic
phase. The nematic-nematic (N$_1$-N$_2$) coexistence region is bounded by a
lower critical point.

Fig.~\ref{fig5} (a), (b): The I-N coexistence regime is bounded by 
an upper critical point above which a single  stable nematic phase is found.
The nematic phase demixes into two nematic phases (N$_1$, N$_2$) at 
sufficiently large values of $\mu_2$.

Fig.~\ref{fig5} (c): The width of the I-N transition broadens 
with increasing aspect ratio $\zeta_2=L_2/D_2$ 
(as compared to Fig.~\ref{fig5} (a), (b)) at fixed $L_1$, $D_1$. 
Moreover, the lower critical point of the N$_1$-N$_2$ coexistence region
shifts to smaller values of $\mu_2$, until the N$_1$-N$_2$ 
and I-N coexistence regions start to overlap, giving rise to a triple point.

A simple estimate of the critical aspect ratios required for a 
depletion-driven demixing transition is obtained from a second virial 
approximation of the excess free energy functional (\ref{eq15}). A 
straightforward analysis shows that demixing at constant volume 
is possible, provided the following interaction parameter $\chi(s_1,s_2)>0$:
\begin{eqnarray}   \label{eq38}
\!\!\!\!\!\chi(s_1,s_2)&=&2E_{12}(s_1,s_2)\!-\!E_{11}(s_1,s_1)\!-\!E_{22}(s_2,s_2),\!
\end{eqnarray}
with 
\begin{eqnarray}   \label{eq39}
\!\!\!\!\!\lefteqn{E_{ij}(s_i,s_j)=}\nonumber
\\&&\frac{1}{3}(D_i+D_j)\left[2(D_iD_j+L_iL_j)(1-s_is_j)\right.\nonumber
\\&&+\left.(D_iL_j+D_jL_i)(1+2s_is_j)+3(D_iL_i+D_jL_j)\right], \label{eq40}
\end{eqnarray}
where $i,j=1,2$.
The values of the nematic order parameters $s_1$ and $s_2$ are obtained
by a numerical minimization of the grand potential. In order study a
possible nematic-nematic demixing transition one may investigate 
the interaction parameter  $\chi(1,1)$:
\begin{eqnarray}   \label{eq41}
\!\!\!\!\!\lefteqn{\frac{\chi(1,1)}{D_1^3}=}\nonumber
\\\!\!\!\!\!&&\left(1+\frac{D_2}{D_1}\right)^2
\left(\zeta_1+\zeta_2\frac{D_2}{D_1}\right)
-8\zeta_1-8\zeta_2
\left(\frac{D_2}{D_1}\right)^3\,,
\end{eqnarray}
where $\zeta_1=L_1/D_1$ and  $\zeta_2=L_2/D_2$. For the model parameters
used in Fig.~\ref{fig5}, $\chi(1,1)>0$, in agreement with that fact 
that nematic-nematic demixing has been found by numerical minimization 
of the FMT grand potential.

Figure 6 displays an alternative representation of the phase diagram shown 
in Fig.~\ref{fig5} (a) in terms of the number densities of the large and the 
small platelets $\rho_1$ and $\rho_2$, respectively. 
\begin{figure}
\vspace*{-1.2cm}
\includegraphics[width=1\linewidth]{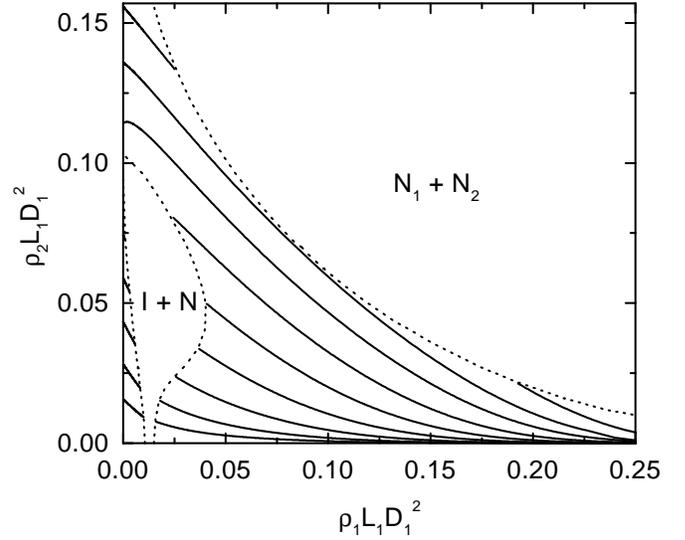}
\vspace*{-4.5cm}
\caption{Phase diagram of a binary platelet fluid, corresponding to
Fig.~\ref{fig5} (a), in the density-density, i.e., $\rho_1$-$\rho_2$  
plane. The dotted lines indicate phase boundaries. The chemical potential 
of the small platelets is kept fixed for each solid line and increases 
from bottom to top: 
$\mu_2/(k_BT)=-5, -4, -3, -2, -1, 0, 1, 1.9, 3$.}
\label{fig6}
\end{figure}
The figure illustrates 
how the composition of the mixture varies upon increasing the chemical 
potential (and hence the reservoir density) of the small platelets.

\section{Conclusions}
We have generalized Zwanzig's model of cuboids (or parallelepipeds) with
only three allowed orientations to the case of monodisperse and bidisperse
systems of platelets ($\zeta=L/D<1$). In view of the uncertainties concerning
the convergence of the virial series, we have used Rosenfeld's FMT, along the
lines of Cuesta's work on parallel hard cubes \cite{cues:96} to derive a free
energy function of the density and the nematic order parameter, which on
the basis of earlier experience with other hard core systems, is expected to be
quantitatively  accurate. In the monodisperse case, the isotropic to nematic 
transition narrows with increasing aspect ratio $\zeta$. Very rich phase
diagrams, involving an isotropic and one or two nematic phases of different
concentrations are found for bidisperse systems; the phase diagrams are 
very sensitive to the size and
aspect ratios and although our results can only be considered to be of
qualitative significance, due to the restricted number of allowed orientations,
we believe that these results clearly point to the possibility of tuning
the phase behaviour by appropriate choices of the relative diameters and
thicknesses of real platelets. This is of genuine technological importance
for the design of clay based drilling fluids \cite{mait:00}. Since clay platelets
are highly charged, Coulombic interactions must be included in the model 
and this could lead to major changes in the predicted phase diagrams.
Calculations of binary platelet phase diagrams, including Coulombic
forces, along the lines of Ref.~\cite{rowa:02} are under way.

\acknowledgements
The authors benefited from helpful discussions with Edo Boek and Geoff 
Maitland. D.G.R. acknowledges the support of EPSRC.

\begin{appendix} 
\section{Third and fourth virial coefficients for the Zwanzig model of platelets}
According to the standard definition of the third virial coefficient 
involving three particles of species $\beta_1$, $\beta_2$, $\beta_3$ 
(where here $\beta_i=x, y$ or $z$ denote one of of the three possible 
orientations of the platelets):
\begin{eqnarray}  \label{eqA1}
\lefteqn{B_3(\beta_1,\beta_2,\beta_3)=}\nonumber
\\&&-\frac{1}{3V}\int d {\bf r}_1\, d {\bf r}_2\, d {\bf r}_3\, 
f_{\beta_1,\beta_2}({\bf r}_{12})
f_{\beta_1,\beta_3}({\bf r}_{13})f_{\beta_2,\beta_3}({\bf r}_{23})
\nonumber
\\=&&-\frac{1}{3}\int d {\bf r}\, d {\bf r}'\, 
f_{\beta_1,\beta_2}({\bf r})f_{\beta_1,\beta_3}({\bf r}')
f_{\beta_2,\beta_3}({\bf r}-{\bf r}')\nonumber
\\=&&\frac{1}{3}\prod_\alpha\int\limits_{-\infty}^\infty d r_\alpha\,
\int\limits_{-\infty}^\infty d r'_\alpha\,
\Theta(S_{\alpha;\beta_1,\beta_2}-|r_\alpha|)\nonumber
\\&&\times\Theta(S_{\alpha;\beta_1,\beta_3}-|r'_\alpha|)
\Theta(S_{\alpha;\beta_2,\beta_3}-|r_\alpha-r'_\alpha|)\,.
\end{eqnarray}
where
$S_{\alpha;\beta_1,\beta_2}=(S_{\alpha;\beta_1}+S_{\alpha;\beta_2})/2$
and definitions (\ref{eq3}) and (\ref{eq4}) have been used. The two-dimensional
integrals involving Heaviside step functions are easily calculated, and 
lead to the following results:
\begin{eqnarray}  \label{eqA2}
B_3(x,x,x)&=&9L^2D^4\,,
\\B_3(x,x,y)&=&LD^3(L+2D)(D+2L)\,,
\\B_3(x,y,z)&=&\frac{1}{3}D^3(2L+D)^3\,,
\end{eqnarray}
while for the isotropic phase (where $\rho_x=\rho_y=\rho_z=\rho/3$), 
$B_3$, as given by (\ref{eq8b}), reduces to:
\begin{eqnarray}  \label{eqA3}
B_3&=&\frac{2}{27}D^6+\frac{16}{9}D^5L+\frac{47}{9}D^4L^2+\frac{52}{27}D^3L^3\,.
\end{eqnarray}
These expressions are easily generalized to the case of binary mixtures 
of platelets, with the added complication of species indices.

Similarly, there are 4 independent 4-th virial coefficients in the 
one-component case, corresponding to various orientations of the four 
particles, namely 
$B_4(x,x,x,x)$, $B_4(x,x,x,y)$, $B_4(x,x,y,y)$ and $B_4(x,x,y,z)$.
Each of these has contributions from 4 irreducible diagrams 
(see Eq.~(4.5.13) in \cite{hans:86})
for each set of orientations. The calculations of the multiple 
integrals are tedious, and we have evaluated $B_4$ only for three 
limiting cases. We find for the isotropic phase:
 
(a) Long rods, corresponding to the limit $\zeta \to \infty$:
\begin{eqnarray}  \label{eqA4}
B_4&=&-\frac{32}{243}D^3L^6\,.
\end{eqnarray}

(b) Thin platelets, corresponding to the limit $\zeta \to 0$:
\begin{eqnarray}  \label{eqA5}
B_4&=&-\frac{16}{243}D^9\,.
\end{eqnarray}

(c) Parallel cubes, corresponding to $\zeta=1$:
\begin{eqnarray}  \label{eqA6}
B_4&=&\frac{34}{3}D^9\,.
\end{eqnarray}

In the case of long rods and thin platelets the fully connected 
diagram does not contribute to $B_4$, and the calculations 
of the other diagrams is considerably simplified.

\end{appendix}

\end{document}